\documentclass[fleqn,twoside]{article}
\usepackage{espcrc2}

\usepackage{graphicx}
\usepackage[figuresright]{rotating}

\newcommand{\AmS}{{\protect\the\textfont2
  A\kern-.1667em\lower.5ex\hbox{M}\kern-.125emS}}

\title{The Paraldor Project}

\author{T. Ashby, D. Galletly, B. Jo\'o, A. D. Kennedy,
G. Lacagnina\\University of Edinburgh}
\begin{document}

\begin{abstract}
Paraldor is an experiment in bringing the power of categorical
languages to lattice QCD computations. Our target language is Aldor,
which allows the capture of the mathematical structure of physics
directly in the structure of the code using the concepts of
categories, domains and their inter-relationships in a way which is
not otherwise possible with current popular languages such as Fortran,
C, C++ or Java.  By writing high level physics code portably in Aldor,
and implementing switchable machine dependent high performance
back-ends in C or assembler, we gain all the power of categorical
languages such as modularity, portability, readability and efficiency.

\vspace{1pc}
\end{abstract}

\maketitle

\section{INTRODUCTION}
The main goal of the Paraldor project is to investigate to what extent
we can use a high level categorical language like Aldor\cite{1} to
write programs for large scale numerical simulations, and in
particular for lattice QCD, on massively parallel computers.  The
design constraints are that we want a code that is portable over many
different architectures e.g., SMP, MPP, Cluster, Workstations, Vector,
Short Vector etc. At the same time we also want efficiency on all
architectures, meaning the code has to interface with assembler
kernels and their data layouts. We do not wish to rely on the compiler
to make clever optimizations, but merely to perform simple
transformations such as inlining. We want modular code where we have
abstracted away both the machine layout and the mathematical structure
from the code so we know only the structure relevant for a particular
algorithm. This should lead to reusability since by partitioning the
problem effectively we leave clear top level code which can easily be
changed without knowledge of, or alteration of low level code.

\section{ALDOR STRUCTURE}
The structure of Aldor is based on categories and domains. Categories
contain signatures expressing only the necessary structure of the
category. Domains are particular instances of a category. They
implement the signatures of the category with perhaps some extra
structure, however the only common structure required by different
domains is that of the category. It is important to notice the
difference between this model and the analogous model of base classes
and derived classes in C++. This is perhaps most clearly illustrated
by the relationships ``domain $\in$ category whereas class $\subset$
base class''. Interfaces in Java use the same model as base classes,
and so don't gain anything over the C++ mechanism.

With the Aldor structure we can write physics code purely in terms of
categories and their signatures. By doing this we ensure that the code
can be used with any domain implementing the category.  This means we
can use our code on different machines or choose a different algorithm
for a particular sub-task without any real change to the code. We can
also build categories with more structure from a tower of simpler
ones. Algorithms can then be written purely in terms of the minimum
categorical structure they need.

\section{COMPARISON}
It is perhaps necessary at this point to justify why this approach
would be useful when there are many other more popular languages in
use. What deficiences are we trying to address and what advantages are
we looking to gain ?

Until recently the standard practice was to write large codes in C or
Fortran. For particularly time critical operations assembler kernels
were added to optimize routines. This of course has the problem that
we cannot write completely modular code. The codes must rely on too
much global data making them very hard to maintain, and the assembler
routines are completely architecture dependent. To write generic
routines we must use macros/templates which have the disadvantages of
lacking type safety and usually having poor error reporting.

When looking at object-oriented languages the obvious benchmarks for
comparison are C++ and Java. What does Aldor have to offer over these
languages ?

Firstly the idea of categories has advantages over inheritance.  In
particular C++ cannot distinguish between things with the same
structure.  An example of this would be if we had classes SU(2) and
SU(3) which were derived from a base class ``group''. Since in C++
these would be seen as subsets of the base class with some extra
structure, the binary operation defined in ``group'' would still be
valid, so we could form nonsensical products like $x*y$ where $x \in$
SU(2) and $y \in $SU(3). This is avoided in the categorical approach
as all that is stored is a signature of the form ``each domain D
implements a product which takes two elements of D and returns an
element of D'', so operations are only allowed within an individual
domain. Java has the same problem, it just requires a cast to convert
an argument of type interface to a specific class. It would assume the
binary operation defined in the first class was the correct one then
attempt to cast the element of the second class into the first. If
this cast was not possible as in the above example it would return a
ClassCastException. Although this catches the error, the cast checking
is done at run time and not at compile time which is clearly much less
desirable.

Aldor is a very strongly typed language. It allows operator
overloading which is forbidden in Java. It has parameterized types and
allows overloading on return types, neither of which are permitted in
Java or C++. These features allow us to write generic code once for
all types that are being used with the types being checked at compile
time. The alternative in say C++ is to either write separate routines
for different types, or to write the routines as macros/templates
which again have the disadvantages of no compile time type checking
and are hard to debug.

\section{PERFORMANCE}

\begin{figure}[h]
\includegraphics[width=7.0cm]{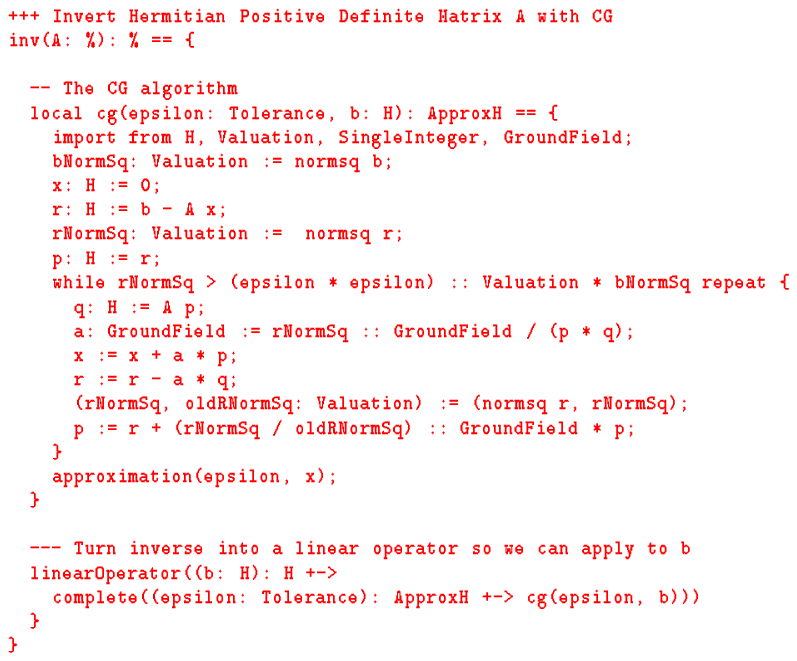}
\caption{\label{code} Aldor Categorical CG code}
\end{figure}

We set up a toy model to test the performance of the code in
comparison to a baseline C code. \\ Consider inverting an $N \times N$
Hermitian positive definite matrix by the Conjugate Gradients (CG)
algorithm. We have implemented the CG algorithm in C, and have
implemented several Aldor variations of the algorithm for comparison.
All the Aldor variations share the same categorical structure (see
Figure \ref{code}), only the back end domains, which implement
matrix-vector and vector-vector operations differ.

We will denote the various code versions as follows:
\begin{description}
\item{C} 
-- Our baseline C Code
\item{A1}
-- Aldor Categories, Aldor Back End, Aldor Garbage Collection (GC)
\item{A2}
-- Aldor Categories, Aldor Back End, Manual Memory Management
\item{A3}
-- Aldor Categories, C Back End, Aldor Garbage Collection
\item{A4}
-- Aldor Categories, C Back End, Manual Memory Management
\end{description}
\begin{figure}[h]
\includegraphics[width=7.0cm]{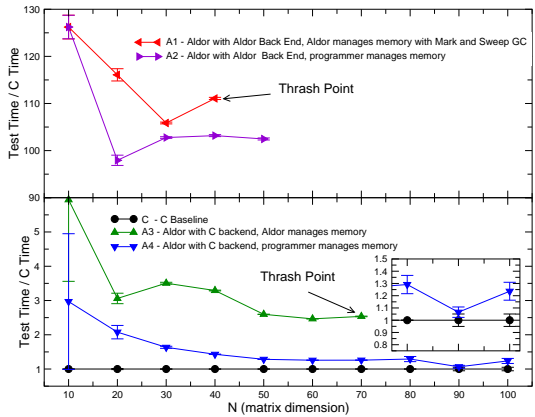}
\caption{\label{ratios} Ratio of test routines to C runtime}
\end{figure}

Our results are shown in Figure \ref{ratios}, where we plot the ratio
of the test run time to the baseline C code runtime for a problem of
the same size. The results from the C code are also plotted.

The Aldor back end test results (A1, A2 - top graph) appear to be over
100 times slower than baseline code and the Aldor mark and sweep
garbage collector is clearly not effective and leads to thrashing. The
C Back End Test Results (A3, A4 - bottom graph) are seen to be of the
same order of magnitude as C Code (until thrashing for A3). For A4 the
ratio of timings is tending towards 1 with increasing problem size. A3
starts to thrash due to memory leak, as the Aldor garbage collector
cannot see memory allocated by the C routines.

From these results a few things become immediately obvious. Firstly
using C-back-ends and memory management, Aldor code performs nearly as
fast as the corresponding C code. Also the overhead decreases with
problem size (and should already be negligible for lattice of $V=2^4$
sites where the fermion matrix has dimension $N=2^4\times 4 \times 3 =
192$). This demonstrates that the complexity of categories etc, is
quite successfully optimised away by the Aldor compiler and that one
can write efficient code using Aldor. It is also clear that memory
management is an important issue. We need ways to manage memory better
than the compiler currently does either by using other garbage
collection schemes or by performing some kind of manual memory
management.
\section{MEMORY MANAGEMENT}

A by-product of writing code in a totally modular way, as was apparent
in the performance results, is that the memory management becomes
highly non trivial. One major problem that arises is that individual
program elements which interact may have no way of knowing when data
is no longer required and hence that memory can be freed. This forces
us to have a more automated memory management scheme. We also have
different types of data to handle. To maintain efficiency we must have
a modular memory management system as well. This leads to the idea of
memory spaces.

With memory spaces we store different types of data in different areas
of memory, each with its own memory management scheme. For large
objects (such as an entire pseudofermion field) a reference counting
method may be preferred as it will free these objects as soon as they
go out of scope with only a small relative overhead in terms of
storage and keeping the reference count. For smaller objects such as
closures it is not as critical that they are cleaned up immediately
and so a mark and sweep process is preferable.
\section{SUMMARY \& FUTURE WORK}
By writing top level physics code in Aldor and implementing switchable
machine dependent back ends in C we have shown in principle that we
can write portable, modular, readable and reusable code without
sacrificing efficiency. The next step is to categorise Krylov space
methods to implement a wide variety of efficient inverters and
eigenvalue solvers. We also have to begin building the physical
objects such as gauge fields and to categorise Monte Carlo algorithms
and other relevant structures with a long term view of a full QCD
code. 
\section{ACKNOWLEDGEMENTS}
This work is supported by the European Community's Human potential
programme under HPRN-CT-2000-00145 Hadrons/LatticeQCD.

\end{document}